\documentclass[]{aastex631}
\usepackage{amsmath}
\usepackage{graphicx}
\usepackage{multirow}

\shorttitle{M67 Sun-like UV}
\shortauthors{Melis \& Kaire}
\begin{document}

\title{\large \bf Ultraviolet variability in Solar-type members of the M67 open cluster}

\author{\large Carl Melis}
\affil{Department of Astronomy \& Astrophysics, University of California San Diego, La Jolla, CA 92093-0424, USA}
\email{cmelis@ucsd.edu}

\author{\large Ekamjot Kaire}
\affil{Department of Astronomy \& Astrophysics, University of California San Diego, La Jolla, CA 92093-0424, USA}



\begin{abstract}

\large{
Solar-type members of {the rich, nearly Solar-age and Solar-metallicity M67 open cluster}
are systematically investigated for ultraviolet variability. 
We utilize archival {\it Galaxy Evolution Explorer} ({\it GALEX}) 
data which features several imaging observation epochs 
spanning 5 years. Stars in or suspected of being in binary systems 
are avoided as well as stars that are blended in {\it GALEX} data, 
leading to a sample of 66 Solar-type stars. 
We assess variability over a variety of timescales 
that probe flares and longer-term trends
that could be due to rotation and activity cycles. 
We do not find conclusive evidence for variability
and determine that Solar-type members of M67 do not display
$>$30\% near-ultraviolet variability over timescales ranging from days to years.
{Furthermore, within 50-second cadence lightcurves generated 
for each of the imaging epochs 
we find no} near-ultraviolet flares that are $\gtrsim$2$\times$ 
the quiescent stellar near-ultraviolet emission level;
the implied ultraviolet flare rate derived from this study is in mild tension with
that derived {for stars observed by {\it GALEX} in the primary {\it Kepler} field.}
This M67 {\it GALEX} study presents one of the most comprehensive ultraviolet 
datasets currently available for probing continuum emission variability for 
old Sun-like stars; the planned NASA UVEX mission has the potential to dramatically 
expand upon this work.
%\\*[1.0mm]
}

\end{abstract}



\large{

\section{\large \bf Introduction}
\label{sec:intro}

Stars are not static objects and experience phenomena ranging from
dramatic and impulsive (flares; e.g., \citealt{kowalski24} and references
therein) to subtle and predictable
(rotation and magnetic cycles; e.g., \citealt{baliunas95}).
With these events come changes in the overall luminous output from the star
that can have different manifestations as a function of wavelength.
While monitoring of the Sun has increasingly covered as many wavelengths as
possible with as few temporal gaps as possible (e.g., \citealt{ayers20}), 
monitoring efforts for other stars have been more sporadic in both temporal and
wavelength coverage.

As of the writing of this article, the stellar and exoplanet communities are
rallying to provide complete characterizations of the luminous output of late-type
stars,
especially within the ultraviolet domain (e.g.,
\citealt{wilson25} and references therein).
Absorption and scattering of stellar light by 
planetary atmospheres can significantly impact the structure, chemistry,
and surface temperature of any orbiting planets 
(e.g., \citealt{krivova09};
\citealt{godolt15}; 
\citealt{france16}; 
\citealt{airapetian20}
and references therein). These processes are strongly wavelength dependent
and ultraviolet light like the 
bright H\,I Ly-$\alpha$ line plays a major role in driving the chemistry and heating
of a planet's middle atmospheric layers, especially through interactions with 
H$_2$O, CH$_4$, and CO$_2$.
Continuum emission in the 1800-2400\,\AA\ range is an important driver for ozone
production while wavelengths up to 3000\,\AA\ heat the middle 
atmosphere.
When trying to identify
Earth-like habitable planets, the interaction of any planet with the 
radiative energy provided by its host star are thus of decisive importance.

While active studies with missions
like {\it TESS} and {\it JWST} will 
imminently inform on the atmospheres and habitability of
planets in the habitable zone around late-type stars {(e.g., \citealt{barkaoui23}; \citealt{madhusudhan25})},
the future is slated to bring a facility
capable of studying habitable zone Earth-like planets around Sun-like stars 
(e.g., \citealt{mamajek24} and references therein).
Through that lens we seek to examine the ultraviolet luminous output from Sun-like
stars, particularly within the temporal domain. Such data has been extensively 
assessed for the Sun 
(e.g., \citealt{floyd03}; \citealt{krivova06}; \citealt{ayers20}) 
and long-term monitoring suggests that while the Sun's total
(integrated over all wavelengths) irradiance changes by about 0.1\%
over a Solar cycle, its irradiance varies by a
few percent between 2000-3000\,\AA , 10-20\% between 1500-2000\,\AA , and more
than 50\% around the Ly-$\alpha$ emission line. 
%The amplitude of variations at 300-400 nm remains rather unclear since it is below the 1-sigma uncertainty of both instruments, which is about 2-3\% (Woods et al. 1996).
A growing collection of studies are exploring the temporal domain of
ultraviolet emission from other {Sun-like} stars with most focusing on line emission 
(e.g., \citealt{dravins93a,dravins93b}; \citealt{dg94}; \citealt{metcalfe07}; \citealt{bm08}; \citealt{ayers20,ayers21}; \citealt{kim22}; \citealt{kamgar24})
and some considering continuum variability
(e.g., \citealt{sofia89}; \citealt{dg94}; \citealt{bm08}; {\citealt{brasseur19};} 
\citealt{ayers20,ayers21}; \citealt{kim22}).
%continuum -- sofia89, dg94 FUV only, ayers20/21 FUV only, bm08+kim22 extract but don't assess
%lines but no time domain? -- olmedo13
%lines -- dg94, dravins93a+b, metcalfe07, ayers20/21, kamgar24

This paper attempts to determine the kind of ultraviolet 
irradiation environment that Solar-type 
stars can provide to any putative Earth-like planets that may orbit around them.
We broadly seek to explore how hospitable Solar-type stars are to potentially habitable
planets; such information will prove invaluable in future investigations of rocky 
planets around other Solar-type stars, especially as it pertains to assessing them 
for currently hosting life.
Specifically, we seek to establish if other Solar-type stars follow the same general
variability patterns as the Sun with a particular focus on continuum emission.
Section \ref{sec:targs} describes the target sample we have selected for this study,
Section \ref{sec:methods} shows the datasets used and how measurements
are made from them,
Section \ref{sec:results} presents our analysis approach and results,
and
Section \ref{sec:conclusions} compares results from this study to
the Sun and appropriate literature datasets with a brief reflection
on how this study can be expanded.

\section{\large \bf Target Sample}
\label{sec:targs}

\begin{figure}
    \centering
    %trim={left bottom right top},clip
    \begin{minipage}{140mm}
     \includegraphics[width=140mm,trim={0.75in 2.9in 0.75in 3in},clip]{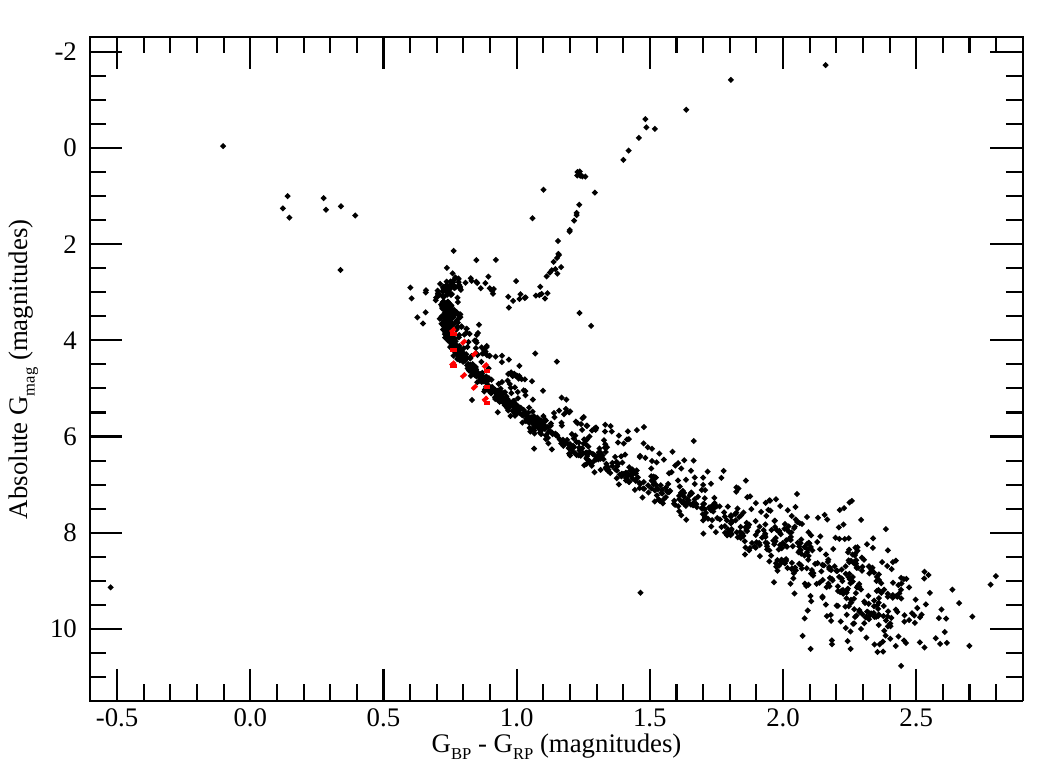}
    \end{minipage}
    \\*
    \begin{minipage}{140mm}
     \includegraphics[width=140mm,trim={0.75in 2.9in 0.75in 3in},clip]{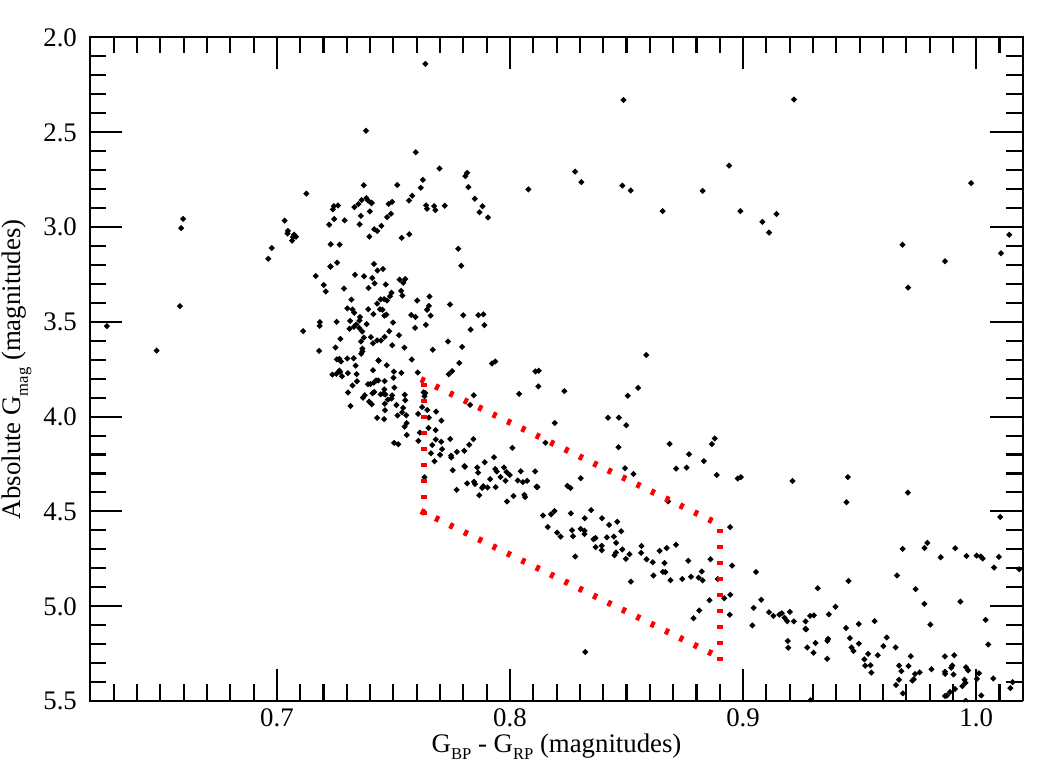}
    \end{minipage}
    \caption{{\it Gaia} DR3 Hertzsprung-Russell diagram for M67 showing the Solar-type target selection; stars selected for this study are enclosed within the red dotted
    lines. No reddening corrections have been applied.
    The top panel shows all cluster members having membership probabililty $>$0.85 as taken from \citet{noormohammadi23}. The bottom panel zooms in on the Solar-type
    star region of the cluster. Error bars are omitted from the plots for the purposes
    of clarity.}
    \label{fig:hr}
\end{figure}

In this study we explore the ultraviolet variability patterns for Solar-type
stars in the M67 open cluster {($d$$=$864\,pc; \citealt{noormohammadi23})}. 
This sample selection is motivated by
several factors. M67 has comparable age (e.g., 
3.95$\substack{+0.16 \\ -0.15}$\,Gyr; \citealt{reyes24}) 
and metallicity ([Fe/H]$\approx$+0.03~dex; \citealt{casamiquela19} and references therein) 
to the Sun which facilitates robust comparison to the Sun's variability properties. 
{\it Gaia} data has resulted in a high-fidelity cluster
membership list (e.g., \citealt{noormohammadi23}; Figure \ref{fig:hr}) 
featuring hundreds of Sun-like stars
while spectroscopic monitoring studies (e.g., 
\citealt{lee08};
\citealt{smolinski11};
\citealt{pasquini12};
\citealt{geller15};
\citealt{brucalassi17};
\citealt{soubiran18};
\citealt{luo19};
\citealt{johnson20};
\citealt{geller21};
\citealt{spina21};
\citealt{abdurrouf22};
\citealt{randich22};
\citealt{rockosi22};
\citealt{hourihane23})
have revealed binary systems that could contaminate activity assessments.
Of utmost importance for the present study is the availability of an
ultraviolet dataset taken with the same instrument spanning a range of temporal
scales (Section \ref{sec:methods}).
The M67 ultraviolet study 
conducted herein provides the first ever look at the ultraviolet continuum
variability levels for a large sample of well-characterized Solar-type stars having 
roughly Solar age and composition.
%Kepler field coverage consisted of a set of three epochs spanning ~1 day (07/03/2010)
% and 4 epochs spanning 72 days centered around 06/28/2009
%so about a year total time baseline and baselines of <~1 day, 1-2 days, ~1 month, ~2 months, and ~1 year

Past works have investigated the activity levels of Sun-like stars within 
M67.
%are any of the Giampapa  stars in my sample? 15
% Cl* NGC 2682 SAND 945   = ID 51
% Cl* NGC 2682 SAND 958   = ID 36
% Cl* NGC 2682 SAND 1048 = ID 48
% Cl* NGC 2682 SAND 1057 = ID 38
% Cl* NGC 2682 SAND 1078 = ID 31
% Cl* NGC 2682 SAND 1087 = ID 29
% Cl* NGC 2682 SAND 1089 = ID 30
% Cl* NGC 2682 SAND 1093 = ID 20
% Cl* NGC 2682 SAND 1095 = ID 49
% Cl* NGC 2682 SAND 1096 = ID 47
% Cl* NGC 2682 SAND 1106 = ID 63
% Cl* NGC 2682 SAND 1218 = ID 50
% Cl* NGC 2682 SAND 1258 = ID 45
% Cl* NGC 2682 SAND 1341 = ID 60
% Cl* NGC 2682 SAND 1449 = ID 34
%which ones show variability?
% 945 -> one low data point?
% 958 -> one of the two 1095s this one (0.62), maybe a down-and-up
% 1048 -> upward trend and one random high point, could be a cycle; nothing in GALEX
% 1057 -> one low data point?
% 1078 -> could be a very weak decaying trend, very low rms; nothing in GALEX
% 1087 -> decaying trend that upturns at end? could be a cycle; GALEX pretty constant
% 1089 -> decaying trend, could be a cycle; GALEX flat 4 data points
% 1093 -> basically stable, one high point?
% 1095 -> rising trend, could be a cycle; GALEX 4 points could be a weak decaying trend
%             => would imply a 20-ish year activity cycle?
% 1096 -> very weak down-then-up, low rms; GALEX flat 4 data points
% 1106 -> basically stable, one high and one low point?
% 1218 -> basically stable, one low point?
% 1258 -> basically stable, one low point?
% 1341 -> basically stable, three barely-in-agreement points
% 1449 -> basically stable, one high and one low point?
\citet{giampapa06} performed Ca\,II H+K activity analysis for 60 Sun-like stars in 
M67 and found that the range of activity levels observed included
stars that were more active than the highest activity levels observed in the Sun.
They also investigated long-term variability in Ca\,II H+K activity for their target 
stars, finding in general that the variability seen in Sun-like stars in M67 has 
higher amplitudes than that for the Sun over a comparable time range 
(Figure 9 from \citealt{giampapa06}).
The reality of especially inactive stars found by \citet{giampapa06} was challenged by 
\citet{curtis17} who explored the impact of interstellar medium (ISM) absorption in removing
Ca\,II H+K flux.
{\citet{curtis17} concludes that ISM absorption varies across the cluster and that any one
cluster star may have more or less ISM absorption (especially from Ca\,II H+K) than another
cluster star. Such} external absorption would not explain variability behavior
or especially active stars, however.
\citet{barnes16} used {\it K2} data to  
derive rotational periods for 20 FGK stars in M67. Their analysis showed that
the Sun appears rather normal in comparison to M67 Solar-like stars in terms 
of rotational properties, thus strengthening our comparison between them.
Despite being covered by three different {\it K2} campaigns
{(497 30-minute cadence lightcurves for 234 stars across the 80-day duration 
campaigns 5, 16, and 18)}, 
\citet{ilin21} only
find one candidate flare event in M67 (originating from a short-orbital period binary 
system as deduced from APOGEE data; \citealt{johnson20} and \citealt{abdurrouf22}) 
suggesting luminous flare events are infrequent or possibly even absent within the 
cluster.

We draw our M67 target sample from \citet{noormohammadi23} and require
prospective objects to have a membership probability $>$0.85. Target stars are
selected from a narrow region of color-absolute magnitude space as
shown in Figure \ref{fig:hr}.
This region is used to obtain stars close to Solar-type, 
to avoid post-main sequence (including possibly early sub-giant) stars, and to ensure
target stars are bright enough to be detected with reasonable significance 
($\gtrsim$10$\sigma$) in at least one band by the 
{\it Galaxy Evolution Explorer} ({\it GALEX}) satellite (Section \ref{sec:methods}).

%binarity exlusion via RUWE and published RV measurements/results
Further target vetting includes systematic searches for evidence of binarity
in both {\it Gaia} and published radial velocity data. Candidates lying appreciably
above the main cluster isochrone were excluded (Figure \ref{fig:hr}). Any target found to
have a renormalized unit weight error (RUWE) $\geq$1.1 is flagged as potentially
in a binary system as well as any star with detected radial velocity variability
of any sort (we also exclude a few stars with no evidence for radial velocity
variability but where the measured value was $>$5$\sigma$ different
from the cluster mean of +33.64\,km\,s$^{-1}$; \citealt{geller15}). 
A final requirement is that target 
stars in the {\it GALEX} images of M67 must be sufficiently
isolated 
%no other detectable GALEX sources within 20"?
to perform accurate photometry; each star was visually inspected in a
{\it GALEX} near-ultraviolet (NUV) image and rejected if we could not construct
a clean photometric extraction aperture and sky annulus (more details are in
Section \ref{sec:methods}).
Furthermore, any star not featured in more than one {\it GALEX} epoch was excluded
from our analysis.
Table \ref{tab:targs} lists all target stars retained for photometric variability
analysis.
%while Table \ref{tab:rejects} reports targets that may have been initially considered but were ultimately rejected.

\section{\large \bf {\it GALEX} photometry and lightcurves} 
\label{sec:methods}

We utilized the multi-epoch visits to M67 by the {\it GALEX} satellite to conduct
our investigation. {\it GALEX} was capable of obtaining 
near-ultraviolet (NUV; 1771-2831\,\AA , $\lambda$$_{\rm eff}$ 2315.7\,\AA ) and 
far-ultraviolet (FUV; 1344-1786\,\AA, $\lambda$$_{\rm eff}$ 1538.6\,\AA ) 
imaging over a $\approx$1.2$^{\circ}$ field-of-view \citep{morrissey07}.
{\it GALEX} was demonstrated to have photometric repeatability between
visits of 0.03~magnitudes in the NUV and 0.05~magnitudes in the FUV
\citep{morrissey07}; these values should be viewed
as noise floors for each filter. 
Table \ref{tab:epochs} lists all epochs used; any collection of visits 
spanning $<$10~hours are treated as a single epoch.
Not all epochs covered every star in the sample and some epochs were taken when
the FUV channel was not operating. Magnitudes reported from {\it GALEX}
are in the AB magnitude system ({\it Gaia} magnitudes are in the Vega system).

\startlongtable
\begin{deluxetable}{lcccccccc}
\label{tab:targs}
\tabletypesize{\small}
\tablewidth{0pt} 
\tablecaption{M67 Solar-type Targets}
\tablehead{ \colhead{ID} &
            \colhead{J2016 RA/DEC} &
            \colhead{J2016 RA/DEC} &
            \colhead{$\pi$} &
            \colhead{G$_{\rm mag}$} &
            \colhead{M$_{\rm G}$} &
            \colhead{G$_{\rm BP}$$-$G$_{\rm RP}$} &
            \colhead{RUWE} &
            \colhead{P$_{\rm memb}$} \\
            \colhead{} &
            \colhead{(sexagesimal)} &
            \colhead{(degrees)} &
            \colhead{(mas)} &
            \colhead{(mag)} &
            \colhead{(mag)} &
            \colhead{(mag)} &
            \colhead{} &
            \colhead{} \\
}
\startdata
  1 &   08 51 58.06  +12 00 59.59 &    132.9919027 +12.01655387 &   1.124 &  13.638 &   3.892 &   0.763 &   0.959 &   0.998 \\
  2 &   08 54 02.68  +12 08 04.20 &     133.5111697 +12.1345012 &   1.139 &  13.682 &   3.965 &   0.764 &   0.957 &   0.978 \\
  3 &   08 50 57.19  +12 05 36.41 &   132.7382901  +12.09344655 &   1.116 &  13.821 &   4.060 &   0.765 &   0.977 &   0.999 \\
  4 &   08 52 09.60  +11 46 46.19 &   133.0400139  +11.77949795 &   1.176 &  13.884 &   4.235 &   0.767 &   0.946 &   0.999 \\
  5 &   08 53 21.72  +11 17 14.62 &    133.3405192  +11.2873942 &   1.150 &  13.767 &   4.070 &   0.768 &   1.027 &   0.986 \\
  6 &   08 50 09.93  +11 59 53.91 &   132.5413914  +11.99830746 &   1.170 &  13.780 &   4.120 &   0.768 &   1.019 &   0.999 \\
  7 &   08 50 40.47  +11 42 11.53 &   132.6686445  +11.70320336 &   1.145 &  13.908 &   4.201 &   0.769 &   1.011 &   0.998 \\
  8 &   08 51 53.18  +12 03 31.98 &    132.9715656  +12.0588830 &   1.129 &  13.758 &   4.021 &   0.770 &   1.035 &   0.997 \\
  9 &   08 51 56.53  +11 48 12.49 &   132.9855281  +11.80346966 &   1.178 &  13.816 &   4.171 &   0.770 &   1.056 &   0.999 \\
 10 &   08 51 48.45  +11 42 22.86 &   132.9518937  +11.70635008 &   1.159 &  13.885 &   4.206 &   0.774 &   1.039 &   0.939 \\
 11 &   08 51 12.31  +11 18 38.72 &   132.8012981  +11.31075419 &   1.151 &  13.909 &   4.215 &   0.774 &   0.982 &   0.996 \\
 12 &   08 50 18.67  +12 16 01.00 &    132.577804  +12.26694538 &   1.146 &  13.986 &   4.283 &   0.775 &   1.074 &   0.997 \\
 13 &   08 51 38.73  +11 42 37.36 &   132.9113613  +11.71037788 &   1.100 &  13.978 &   4.185 &   0.777 &   0.975 &   0.999 \\
 14 &   08 53 40.02  +12 05 01.43 &    133.4167436  +12.0837308 &   1.125 &  13.923 &   4.180 &   0.780 &   0.953 &   0.992 \\
 15 &   08 51 15.04  +11 49 21.09 &    132.8126575  +11.8225261 &   1.136 &  13.987 &   4.264 &   0.780 &   1.014 &   0.999 \\
 16 &   08 50 38.66  +11 12 12.84 &    132.6610991 +11.20356562 &   1.120 &  14.015 &   4.261 &   0.780 &   0.971 &   0.998 \\
 17 &   08 50 56.51  +11 38 08.17 &   132.7354783  +11.63560363 &   1.186 &  13.981 &   4.352 &   0.781 &   0.979 &   0.999 \\
 18 &   08 53 34.71  +11 41 58.50 &    133.3946156  +11.6995826 &   1.135 &  13.662 &   3.938 &   0.782 &   0.987 &   0.995 \\
 19 &   08 52 30.26  +11 42 50.42 &   133.1261032  +11.71400676 &   1.068 &  13.976 &   4.119 &   0.784 &   0.988 &   0.997 \\
 20 &   08 51 18.71  +11 55 49.68 &    132.8279619  +11.9304678 &   1.142 &  13.979 &   4.268 &   0.785 &   1.018 &   0.999 \\
 21 &   08 50 56.31  +11 51 29.24 &   132.7346429  +11.85812287 &   1.167 &  13.959 &   4.295 &   0.786 &   0.972 &   0.997 \\
 22 &   08 50 11.60  +12 14 33.62 &   132.5483503  +12.24267153 &   1.166 &  14.034 &   4.367 &   0.788 &   0.935 &   0.994 \\
 23 &   08 50 00.83  +11 42 09.79 &   132.5034644  +11.70271836 &   1.176 &  14.022 &   4.374 &   0.790 &   1.004 &   0.997 \\
 24 &   08 50 48.10  +11 54 47.66 &   132.7004148  +11.91323806 &   1.128 &  13.952 &   4.214 &   0.793 &   0.974 &   0.999 \\
 25 &   08 50 22.79  +12 13 37.12 &   132.5949661  +12.22697823 &   1.114 &  14.041 &   4.276 &   0.793 &   0.983 &   0.997 \\
 26 &   08 52 11.38  +11 40 32.04 &   133.0474329  +11.67556711 &   1.147 &  14.074 &   4.371 &   0.793 &   1.020 &   0.998 \\
 27 &   08 52 09.93  +12 10 31.05 &    133.0413877  +12.1752927 &   1.141 &  14.002 &   4.289 &   0.794 &   0.891 &   0.998 \\
 28 &   08 51 34.11  +11 46 55.74 &    132.892119  +11.78215118 &   1.143 &  14.028 &   4.318 &   0.795 &   0.935 &   0.998 \\
 29 &   08 51 23.13  +11 54 04.98 &    132.8463869 +11.90138407 &   1.162 &  14.013 &   4.338 &   0.798 &   0.971 &   0.999 \\
 30 &   08 51 12.28  +11 54 22.96 &   132.8011626  +11.90637877 &   1.166 &  14.004 &   4.337 &   0.803 &   0.966 &   0.999 \\
 31 &   08 51 27.48  +11 53 02.43 &   132.8645079  +11.88400937 &   1.162 &  14.019 &   4.345 &   0.805 &   0.970 &   0.999 \\
 32 &   08 50 28.04  +11 54 50.54 &   132.6168447  +11.91403833 &   1.174 &  14.022 &   4.370 &   0.811 &   0.932 &   0.857 \\
 33 &   08 52 01.55  +12 03 54.19 &   133.0064639  +12.06505174 &   1.159 &  14.051 &   4.371 &   0.811 &   1.003 &   0.996 \\
 34 &   08 51 56.57  +11 47 24.96 &    132.985721  +11.79026558 &   1.148 &  14.223 &   4.522 &   0.814 &   1.055 &   0.998 \\
 35 &   08 51 55.14  +11 39 40.16 &   132.9797563  +11.66115531 &   1.152 &  14.207 &   4.515 &   0.817 &   1.010 &   0.999 \\
 36 &   08 51 14.33  +11 41 09.18 &   132.8096959  +11.68588288 &   1.139 &  14.329 &   4.611 &   0.820 &   0.969 &   0.999 \\
 37 &   08 50 09.58  +11 38 41.91 &   132.5399361  +11.64497559 &   1.148 &  14.332 &   4.632 &   0.821 &   1.008 &   0.997 \\
 38 &   08 51 20.03  +11 51 01.65 &   132.8334417  +11.85045922 &   1.112 &  14.146 &   4.377 &   0.825 &   1.012 &   0.998 \\
 39 &   08 51 36.02  +12 23 47.11 &   132.9001008  +12.39641961 &   1.150 &  14.206 &   4.509 &   0.826 &   0.928 &   0.987 \\
 40 &   08 50 21.56  +11 50 23.12 &   132.5898482  +11.83975787 &   1.141 &  14.345 &   4.630 &   0.826 &   1.084 &   0.998 \\
 41 &   08 50 58.33  +11 58 14.79 &   132.7430592  +11.97077596 &   1.124 &  14.337 &   4.592 &   0.830 &   0.998 &   0.984 \\
 42 &   08 51 53.88  +11 39 04.81 &   132.9744811  +11.65133687 &   1.132 &  14.349 &   4.619 &   0.831 &   0.949 &   0.999 \\
 43 &   08 52 00.40  +11 56 07.01 &   133.0016787  +11.93527943 &   1.169 &  14.196 &   4.536 &   0.832 &   0.937 &   0.999 \\
 44 &   08 51 12.52  +12 11 17.30 &   132.8021731  +12.18814015 &   1.132 &  14.222 &   4.492 &   0.834 &   0.985 &   0.998 \\
 45 &   08 51 48.93  +11 48 02.16 &    132.953887  +11.80060118 &   1.175 &  14.330 &   4.680 &   0.839 &   1.006 &   0.999 \\
 46 &   08 52 15.44  +11 53 01.46 &    133.0643474  +11.8837387 &   1.182 &  14.340 &   4.703 &   0.839 &   1.000 &   0.998 \\
 47 &   08 51 12.97  +11 57 00.88 &   132.8040507  +11.95024464 &   1.176 &  14.285 &   4.636 &   0.841 &   0.988 &   0.998 \\
 48 &   08 51 27.95  +11 50 11.87 &   132.8663975  +11.83663088 &   1.168 &  14.234 &   4.571 &   0.842 &   1.063 &   0.999 \\
 49 &   08 51 09.13  +11 57 00.31 &   132.7880497  +11.95008719 &   1.118 &  14.391 &   4.632 &   0.844 &   0.933 &   0.998 \\
 50 &   08 51 34.49  +11 43 49.44 &   132.8936995  +11.73040139 &   1.154 &  14.419 &   4.730 &   0.844 &   1.013 &   0.999 \\
 51 &   08 51 23.86  +11 38 52.11 &   132.8494072  +11.64780855 &   1.173 &  14.370 &   4.716 &   0.845 &   0.947 &   0.999 \\
 52 &   08 49 36.68  +12 30 53.41 &    132.4028389 +12.51483483 &   1.107 &  14.445 &   4.665 &   0.845 &   0.979 &   0.962 \\
 53 &   08 54 10.58  +12 16 37.62 &   133.5440834 +12.277117990 &   1.132 &  14.499 &   4.767 &   0.861 &   1.036 &   0.930 \\
 54 &   08 51 30.08  +11 27 31.50 &     132.8753246  +11.458749 &   1.153 &  14.441 &   4.750 &   0.849 &   0.974 &   0.998 \\
 55 &   08 52 45.49  +11 54 45.70 &   133.1895473  +11.91269295 &   1.133 &  14.454 &   4.726 &   0.851 &   0.930 &   0.998 \\
 56 &   08 50 52.29  +11 45 03.55 &   132.7178854  +11.75098493 &   1.205 &  14.465 &   4.870 &   0.851 &   0.979 &   0.999 \\
 57 &   08 51 01.46  +12 28 45.30 &   132.7560722 +12.479248025 &   1.171 &  14.496 &   4.838 &   0.861 &   1.091 &   0.979 \\
 58 &   08 51 35.46  +11 34 32.12 &   132.8977629  +11.57558909 &   1.161 &  14.496 &   4.819 &   0.865 &   0.984 &   0.998 \\
 59 &   08 53 12.28  +11 40 28.43 &   133.3011769  +11.67456306 &   1.181 &  14.501 &   4.862 &   0.868 &   1.007 &   0.996 \\
 60 &   08 51 30.74  +12 04 16.03 &    132.878095  +12.07112058 &   1.109 &  14.535 &   4.759 &   0.876 &   0.992 &   0.998 \\
 61 &   08 52 47.65  +11 57 26.16 &   133.1985207  +11.95726601 &   1.103 &  14.631 &   4.844 &   0.877 &   0.953 &   0.998 \\
 62 &   08 51 09.28  +11 33 32.98 &   132.7886738  +11.55915992 &   1.223 &  14.627 &   5.063 &   0.878 &   1.008 &   0.999 \\
 63 &   08 51 07.94  +11 58 53.32 &   132.7830938  +11.98147888 &   1.136 &  14.572 &   4.849 &   0.880 &   1.064 &   0.997 \\
 64 &   08 51 59.00  +11 41 49.16 &   132.9958397  +11.69698846 &   1.244 &  14.549 &   5.022 &   0.881 &   0.964 &   0.999 \\
 65 &   08 51 15.83  +11 40 30.28 &     132.8159657 +11.6750777 &   1.122 &  14.613 &   4.864 &   0.882 &   1.012 &   0.999 \\
 66 &   08 50 18.38  +11 54 23.98 &   132.5765685  +11.90666029 &   1.174 &  14.620 &   4.968 &   0.885 &   0.987 &   0.976 \\
\enddata
\tablecomments{RA/DEC, $\pi$, apparent magnitudes (G$_{\rm mag}$, G$_{\rm BP}$$-$G$_{\rm RP}$) and RUWE come from {\it Gaia} DR3 \citep{gaiadr3}. Uncertainties on $\pi$ are
generally 0.02\,mas ($\sim$2\%), but may be larger if astrometric excess noise is present
(this is only the case for a few stars in the sample and never exceeds 0.05\,mas).
Photometric errors are $<$1\%. M$_{\rm G}$ is calculated
from the reported $\pi$ and G$_{\rm mag}$. P$_{\rm memb}$ is the membership probability
as assessed by \citet{noormohammadi23}.}
\end{deluxetable}

To generate photometric measurements in the NUV and FUV (when present),
we employ the {\sf gPhoton} package \citep{million16}.
Table \ref{tab:epochs} lists start and end times and durations for each epoch.
Values are listed in {\it GALEX} time which is 315964800~seconds less than
{\sf UNIX} time (e.g., \citealt{million16}); for reference,
a {\sf UNIX} time of 1204545600.0~seconds (a {\it GALEX} time of 888580800.0~seconds)
corresponds to 12:00 UT 03 March 2008. 
Start and end times as reported
in Table \ref{tab:epochs} are designed to encompass all possible times when
{\it GALEX} may have been recording photons for any given M67 star during a
pointing toward the general direction of the cluster. 
We did not adjust the default
{\sf gPhoton} edge limit that effectively truncates the {\it GALEX} field-of-view
to $\approx$1.1$^{\circ}$.
Table \ref{tab:epochs} start and end times are
fed directly into the {\sf gPhoton} gAperture module when producing epochal
photometry. Epoch start and end times are derived by examining
the range of start (end) times for all stars in any given epoch as returned
from the {\sf gPhoton} gFind module, taking the earliest (latest) value, and
padding it with $-$2000 (+2000) seconds. The epoch duration listed in 
Table \ref{tab:epochs} is calculated for this full range. The Total Exposure
column of Table \ref{tab:epochs} gives a closer accounting of the actual amount
of exposure time any star may have accrued during each epoch.

Each target is individually inspected in a {\it GALEX} image
of M67 to determine the optimal aperture and background annulus radii
to use for photometric extraction in all epochs; background count levels
are assessed for each individual epoch. Generally,
a radius of 12.8$'$$'$ and annulus spanning 16.0-22.0$'$$'$ are used;
roughly 20\% of the targets benefited from truncated values of 9.0$'$$'$ and
12.8-18.8$'$$'$ respectively when there was potential for
contamination by nearby sources. Aperture sizes are chosen to ensure
source photometry can be reliably compared across all epochs regardless
of where the source fell in the {\it GALEX} field-of-view in any individual
epoch ({\it GALEX} featured a significantly
position-dependent point-spread function; \citealt{morrissey07}).
gAperture is called with source positions as given in Table \ref{tab:targs},
radius and background annulus values as reported in Table \ref{tab:phot}, 
each epoch time range listed in Table \ref{tab:epochs},
and directed to produce a single integrated magnitude and associated uncertainty
for the band of interest (NUV and FUV are run separately).
Table \ref{tab:phot} reports for each source
aperture corrections as suggested in Figure 4 of \citet{morrissey07} and measured
epochal NUV photometry. Figure \ref{fig:nuvhr} plots 
absolute NUV magnitudes versus {\it Gaia} colors
for all Table \ref{tab:targs} targets.

After careful consideration of epochal FUV results, it was determined
that most individual epochs did not actually result in a detection.
Indeed, many of them reported either negative flux values or flux values
smaller than (or certainly $<$2$\times$) the reported flux uncertainty.
We conclude that {\it GALEX} was not sensitive enough to provide epochal
FUV detections for M67 Sun-like stars. However, for the sake of completeness,
we ran the gAperture module for each source covering all time ranges
(encompassing the Epoch 1 start time to the Epoch 12 end time)
to produce an all-epoch integrated FUV magnitude. Even in this situation,
with most sources accruing $\approx$7,000~seconds of total FUV exposure time,
many sources still produced non-detections. 
We converted non-detections to limits by taking 2.5$\times$ the flux uncertainty
as the limiting flux and converting it to an FUV magnitude (any source with
flux $\geq$2.5$\times$ the flux uncertainty was considered a detection).
Table \ref{tab:phot} reports
FUV all-epoch integrated magnitudes or limits and Figure \ref{fig:nuvhr}
plots the recovered values (including limits) as a function of target star
{\it Gaia} color.

Lightcurves are generated with
the same gAperture command used to produce epochal photometry 
for each target with a revised time range
that encompassed the Epoch 1 start time to the Epoch 12 end time.
An additional parameter specified the output as photometry for time bins of 50~seconds
throughout the input time range{; this cadence was motivated by comparison
with \citet{brasseur19} as described in detail in Section \ref{sec:conclusions}}.
Given the inability to detect sources in epochal FUV photometry as described
above, lightcurves are only generated for NUV data.

\begin{deluxetable}{llccccc}
\label{tab:epochs}
\tabletypesize{\small}
\tablewidth{0pt} 
\tablecaption{{\it GALEX} M67 Epochs}
\tablehead{ \colhead{Epoch} &
            \colhead{UT Date} &
            \colhead{MJD} &
            \colhead{Epoch Start} &
            \colhead{Epoch End} &
            \colhead{Epoch Duration} &
            \colhead{Total Exposure} \\
            \colhead{} &
            \colhead{} &
            \colhead{(days)} &
            \colhead{({\it GALEX} time, sec)} &
            \colhead{({\it GALEX} time, sec)} &
            \colhead{(hrs)} &
            \colhead{Time (sec)} \\
}
\startdata 
1  & 2005-01-30 & 53400.58 & 791123761.995 & 791133774.995 & 2.8 & 95 + 96 \\
2  & 2005-02-18 & 53419.59 & 792768195.995 & 792773889.995 & 1.6 & 1694 \\
3  & 2006-01-21 & 53756.35 & 821838478.995 & 821873536.995 & 9.8 & 74 + 1313 + 1458 \\
4  & 2006-01-23 & 53758.38 & 822039744.995 & 822045111.995 & 1.5 & 1367 \\
5  & 2006-01-31 & 53766.73 & 822761930.995 & 822767294.995 & 1.5 & 1364 \\
6  & 2006-02-15 & 53781.72 & 824057421.995 & 824061531.995 & 1.2 & 110 \\ 
7  & 2008-01-28 & 54493.47 & 885552777.995 & 885556885.995 & 1.2 & 108 \\
8  & 2008-03-01 & 54526.69 & 888422119.995 & 888427600.995 & 1.6 & 1481 \\
9  & 2009-01-03 & 54834.48 & 915015145.995 & 915019708.995 & 1.3 & 183 \\
10 & 2009-01-27 & 54858.85 & 917121681.995 & 917125955.995 & 1.2 & 274 \\
11 & 2010-01-12 & 55208.24 & 947308128.995 & 947313806.995 & 1.6 & 1678 \\
12 & 2010-01-13 & 55209.20 & 947390910.995 & 947396613.995 & 1.6 & 1703 \\
\enddata
\tablecomments{See Section \ref{sec:methods} for a description of start, end, and duration times. Total Exposure Time gives the {\it GALEX}-reported on-sky time for all pointings conducted within a given epoch and doesn't account for detector dead time or shuttered time (see \citealt{million16,million23} and references therein); when multiple values appear, this means individual exposures separated by some gap are present within the reported epoch duration. MJD is the median Modified Julian Date value for all
sources with data in a given epoch (individual values are reported in Table \ref{tab:phot}).}
\end{deluxetable}
%some objects have data in NUV lightcurve but do not produce anything in the same time range for epochal photometry; e.g.; obj  6: 821838478.995 - 824061531.995
% looks like that range is bad data and gFind and gAperture with coadd=True ignore it while lightcurve spits out data with a bunch of flags

\begin{figure}[!t]
    \centering
    %trim={left bottom right top},clip
    \begin{minipage}{140mm}
     \includegraphics[width=140mm,trim={0.65in 2.9in 0.75in 3in},clip]{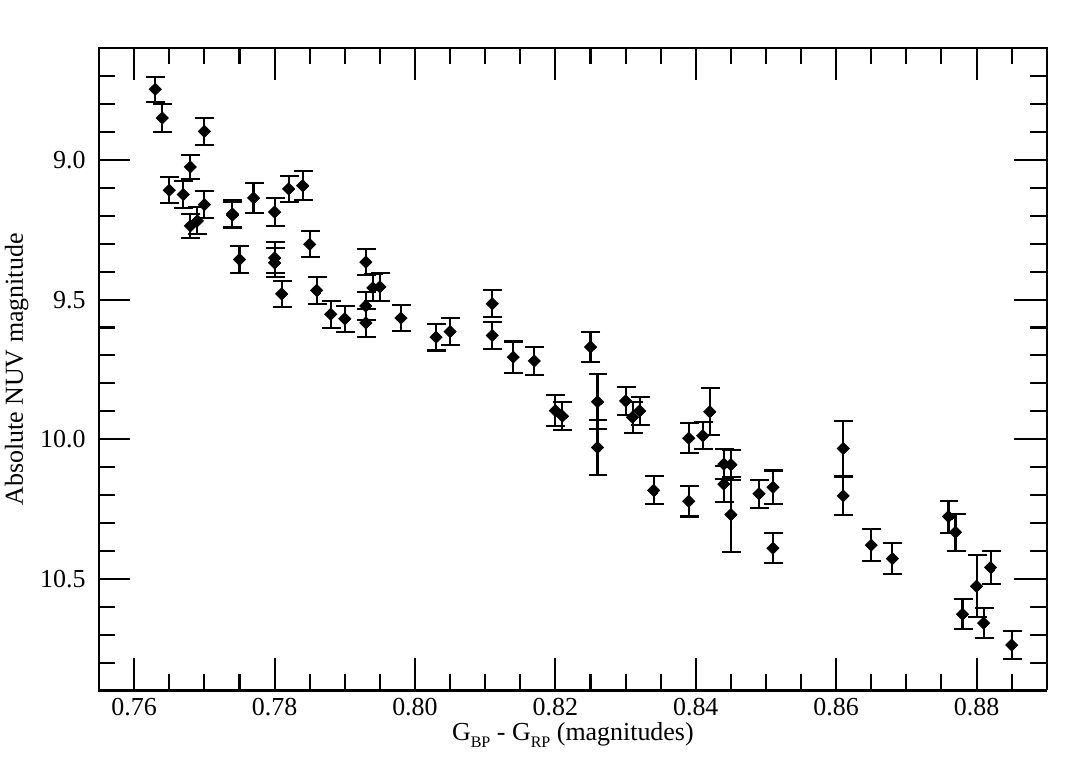}
    \end{minipage}
    \\*
    \begin{minipage}{140mm}
     \includegraphics[width=140mm,trim={0.65in 2.9in 0.75in 3in},clip]{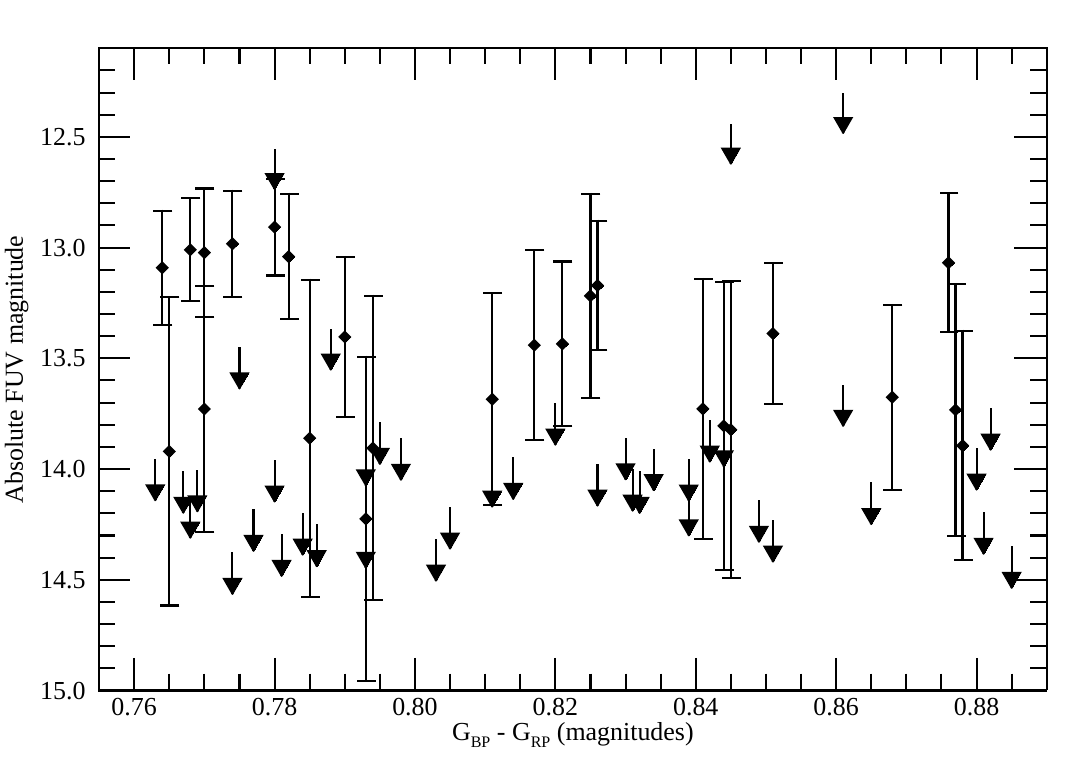}
    \end{minipage}
    \caption{{\it GALEX} and {\it Gaia} DR3 Hertzsprung-Russell diagrams for 
    Table \ref{tab:targs} M67 Solar-type targets. Aperture corrections as listed
    for each source in Table \ref{tab:phot} have been applied.
    Data points are not corrected for reddening (\citealt{taylor07} reports
    an $E$($B$$-$$V$) for M67 of 41$\pm$4\,mmag; see
    \citealt{curtis17} 
    and references therein for a discussion of intervening interstellar medium material
    on an individual target level).
    The absolute NUV magnitude is calculated from the weighted mean of all epochs
    reported in Table \ref{tab:phot}
    for each target and the {\it Gaia} parallax reported in Table \ref{tab:targs}.
    Errors for the absolute NUV magnitude include both the uncertainty
    on the weighted mean (with a noise floor
    of 0.03~magnitudes, see Section \ref{sec:methods}) and the {\it Gaia} parallax uncertainty; absolute NUV magnitude error bars are typically 
    around 0.05~magnitudes. 
    In general, a spread of roughly 0.3-0.4 magnitudes is seen at each
    G$_{\rm BP}$$-$G$_{\rm RP}$ color.
    The absolute FUV magnitude is calculated from the integrated FUV magnitude
    reported in Table \ref{tab:phot} and the {\it Gaia} parallax; 
    see Section \ref{sec:methods} for
    a description of FUV values and associated uncertainties or limits.
    Errors on the
    {\it Gaia} color are $<$1\% and not plotted to maintain clarity.}
    \label{fig:nuvhr}
\end{figure}

\section{\large \bf Analysis and Results} 
\label{sec:results}

Being wary of the potential for varying interstellar medium levels between stars 
within M67 (see \citealt{curtis17} and references therein), we choose to initially focus our analysis 
at an intra-object level. Epochal measurements are assessed for each source,
examining the difference between magnitudes measured in any given pair
of epochs without repeating any epoch pairs (i.e., Epoch 1 $-$ Epoch 2 excludes 
Epoch 2 $-$ Epoch 1; the absolute value of each difference is taken to
remove any effects stemming from which epoch is subtracted).
The significance of magnitude differences are derived by
dividing magnitude differences by the propagated associated uncertainties
for both epochs. In this way we produce Figure \ref{fig:nuvdif} which plots
all valid magnitude differences against their significances.

Figure \ref{fig:nuvdif} exhibits a few interesting features.
The linear, rising bottom envelope for the data points is set by the
significance necessary to detect a magnitude difference given the
error distribution in Table \ref{tab:phot} (e.g., the smallest 
1$\sigma$ is 0.042 magnitudes given the 0.03 magnitude-per-epoch noise floor;
see Section \ref{sec:methods}). Of more interest to the present
study is the collection of epoch differences that have
significance $>$3$\sigma$. Two stars populate the differences
with significance $>$5$\sigma$ while 13 stars total (including the two
with $>$5$\sigma$ differences) contribute to significances $>$3$\sigma$.
While the bottom envelope of significances for 3 and 5$\sigma$ are
roughly 0.13 and 0.21 magnitudes respectively, some stars appear to have
differences up to 0.3~magnitudes.

Appendix \ref{app:A} showcases the epochal lightcurves for a sample of
stars with significant differences
as a sanity check of whether or not their apparent variability is evidence of
activity cycles (the most likely variability mechanism to cause changes
over the epoch timescales from Table \ref{tab:epochs}). 
The plots demonstrate that for the two stars with $>$5$\sigma$ differences
the magnitude changes can be traced back to one discrepant epoch.
However, in the lingering
$>$3$\sigma$ difference sample and evident in the 2-3$\sigma$ population
is a set of stars which present tentative (weak) evidence for possible
activity cycle variations at the 0.1-0.3 magnitude level.

{The weak evidence for variability at an intra-object level makes one wonder
if the vertical spread of magnitudes seen in the color-absolute NUV magnitude plot shown in
the top panel of Figure \ref{fig:nuvhr} could be evidence for variability at an inter-object
level since the spread is comparable to the possible cycle variability discussed in the previous
paragraph.
\citet{curtis17} provides evidence for
variable interstellar Ca~II absorption across the M67 cluster, but it is not clear 
if ISM structure could produce the 
the 0.3-0.4 magnitude spread in absolute NUV magnitudes seen
for each color in Figure \ref{fig:nuvhr}.
Spreads of $\approx$0.2~magnitudes are seen in the absolute G$_{\rm mag}$ distribution
(e.g., Figure \ref{fig:hr}) which seems to argue against both of ISM and activity as the cause
(activity-related variability for the Sun in the visible is $<$1\% or $<$0.01~magnitudes and
NUV and visible extinctions should not be of comparable magnitude).
The color-absolute FUV magnitude plot shown in the bottom panel} of Figure \ref{fig:nuvhr}
is unable to add much to this discussion given the generally low detection
rate and poor signal-to-noise when detections were possibly present.
However, one curious feature of the color-absolute FUV diagram is the apparent
dichotomy between what appear to be robust detections (sources with
absolute FUV magnitudes near 13) and non-detections (sources with limits
between 14-14.5 magnitudes) for the bluest Sun-like M67 stars.
Taken at face value this could possibly be suggestive of stars with
similar physical properties having a spread of greater than a factor of 3 in
FUV flux or it could indicate dramatic local reddening variations within the cluster.
While tantalizing, the low quality of the FUV detections within the M67 Sun-like
star population prevents any conclusive analysis.
Thus, while both Figure \ref{fig:nuvhr} and epochal lightcurve properties
for individual M67 Sun-like stars seem to be suggestive of NUV variability
at the $\sim$0.3 magnitude level, the available data are unfortunately
insufficient to conclusively identify it.

\begin{figure}
    \centering
    %trim={left bottom right top},clip
     \includegraphics[width=140mm,trim={0.60in 2.85in 0.75in 3in},clip]{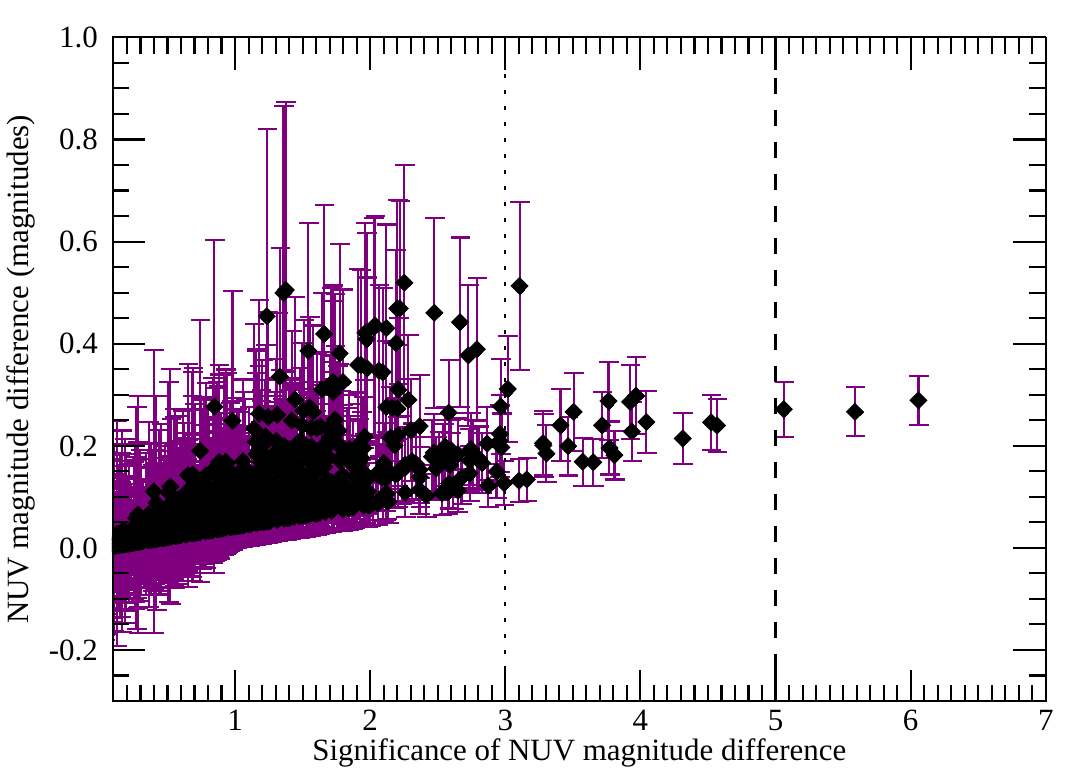}
    \caption{NUV magnitude differences between pairs of epochs for a given star
    plotted as a function of the significance of that magnitude difference 
    (calculated as the
    magnitude difference value divided by the propagated uncertainty for the two 
    epochs); results for all epoch pairs for all stars in the sample are plotted. 
    Black diamonds are the magnitude difference values while the
    purple error bars show the propagated uncertainty; dotted and dashed lines
    guide the eye for assessing $>$3$\sigma$ and $>$5$\sigma$ significance levels
    respectively. A collection of epoch pairs
    appear to show $>$3$\sigma$ significance at the 0.1-0.3 magnitude level;
    see detailed discussion in Section \ref{sec:results}.}
    \label{fig:nuvdif}
\end{figure}

Moving from long-term to short-term variability, we search for evidence of
flares in {\it GALEX} data for our M67 target stars using the 50-second
cadence lightcurves. For each object we
calculate the median magnitude and median uncertainty from all 
acceptable lightcurve samples;
we exclude samples which had $<$25~seconds of integration time
and those which featured any data quality flags as output by {\sf gPhoton}.

We first performed an automated flare search which requires two contiguous
data points to be brighter than the median magnitude by
$>$2$\times$ the median uncertainty for a given star. This returned no
positive hits indicating no flares in the dataset as we had defined them.
We then performed a visual inspection of all flare lightcurves
to ensure this result was robust.
The visual inspection returned the same result and demonstrated that
all lightcurves generally featured well-behaved magnitude distributions that were
consistently contained within the $\pm$2$\times$ median uncertainty bounds
(see Appendix \ref{app:A}).

Using the 2$\times$ median uncertainty as our detection threshold,
we assessed the overall flare sensitivity for each M67 star in our sample.
Figure \ref{fig:lcsens} illustrates this sensitivity and in particular how
it varies as a function of the target star median NUV magnitude.
The range in sensitivities spans from flares that are roughly 1.3$\times$
the stellar NUV flux output on the bright end to flares that are roughly 3$\times$
the stellar NUV flux output on the faint end.

\section{\large \bf Discussion and Conclusions}
\label{sec:conclusions}

We have presented a comprehensive search for ultraviolet variability in the
{\it GALEX} M67 Sun-like star population. Our efforts have failed to
identify clear evidence for variability at any observed timescale, but are possibly
suggestive of long-term changes likely related to activity cycles if real.

\begin{figure}
    \centering
    %trim={left bottom right top},clip
     \includegraphics[width=140mm,trim={0.65in 2.85in 0.75in 3in},clip]{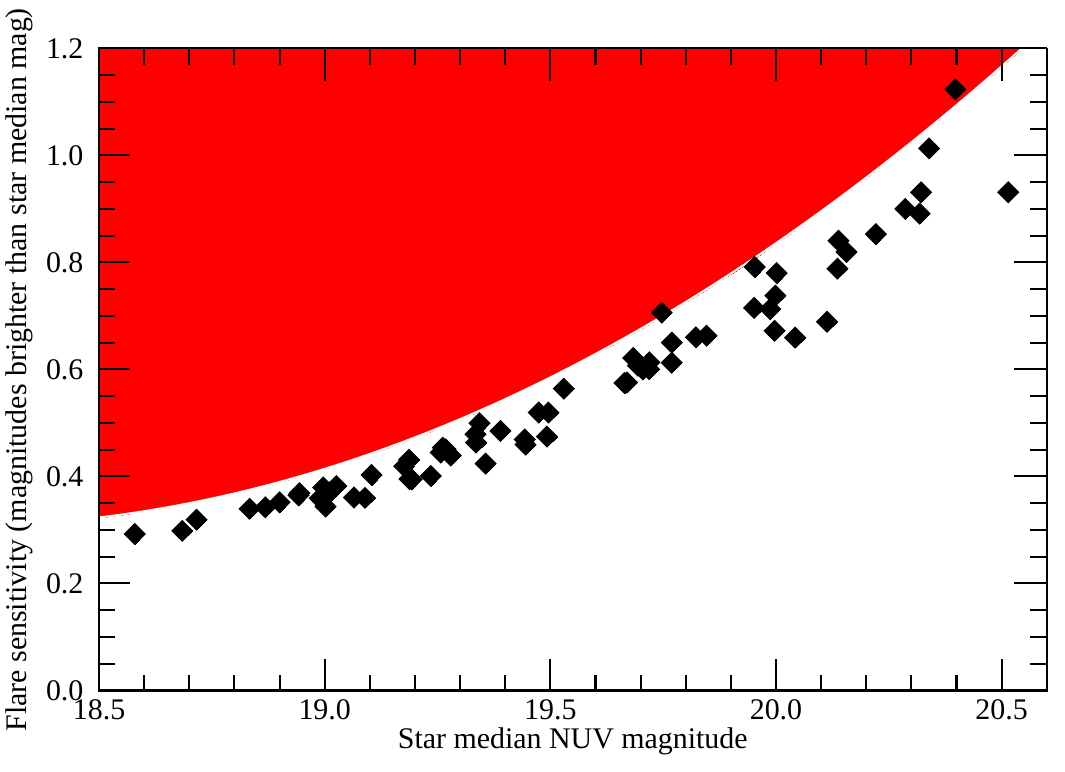}
    \caption{NUV flare sensitivity as derived from lightcurve analysis
             described in Section \ref{sec:results}{; flare sensitivity is defined
             to be 2$\times$ the median uncertainty in the lightcurve for each object}. The red filled region
             denotes where our sensitivity allowed us to detect flares
             but none were seen. Each data point corresponds to a single
             analyzed star; uncertainties on the median NUV magnitude are
             omitted for clarity and are generally in the range of 0.03-0.05
             magnitudes.}
    \label{fig:lcsens}
\end{figure}

%epochal comparison to the Sun and alpha Cen (IUE/HST) + other "trenches" stars
% also mention Giampapa CaII results
For context we consider the continuum variability of the Sun in the 
near-ultraviolet. As stated previously, the Sun can experience activity cycle
variability
{on the order of a few percent (roughly 0.03~magnitudes) from
2000-3000\,\AA\
and 10-20\% (roughly 0.1-0.2~magnitudes) from 1500-2000\,\AA ; Figure 13(b) of
\citet{ayers20} illustrates this well.
Recall the {\it GALEX} NUV bandpass spanned 1771-2831\,\AA ,
so it would have mostly sampled the 2000-3000\,\AA\ region and with our assessed
long-term variability sensitivity limit of $\gtrsim$0.13~magnitudes (see above) would have been unlikely
to detect the roughly few percent Solar-like activity cycle variability amplitudes.
If M67 Sun-like stars indeed experience activity cycle variability at the level 
of 0.3 magnitudes as discussed above, then they would be quite distinct 
from the Sun in activity properties
(roughly ten times the Solar activity cycle variability amplitude in the NUV).}
The sunspot cycle has a period of $\approx$11~years and {\it GALEX}
M67 monitoring epochs feature close to yearly measurements over $\approx$5~years.
If M67 stars have activity cycle periods similar to the Sun
{(and have the necessary variability amplitudes much stronger than the Sun so that 
{\it GALEX} could
detect them)}, then 
with a sample of over 60 objects (and assuming random relative phases for their
activity cycles across the sample) we would be likely to catch at least some
progressing from their activity maximum (or minimum) to their activity minimum
(or maximum) over a 5~year time span resulting in the largest possible change
in flux. No such changes are conclusively detected and whether this is due to longer
activity cycle periods or lower amplitudes (or both) 
remains unknown (but see more about M67 star activity
cycles below). 
%In any case, NUV continuum variability due to activity cycles at the level of the Sun would be undetectable for solar-type stars in M67 with the data set presented herein due to the sensitivity  per epoch afforded by {\it GALEX}.

%Sun cycle min to max flux change
%c2796: min = -5.63 -> 2.3442288
%       max = -5.60 -> 2.5118864
%                 => 6.9% (this is flux under MgII k line!)
%
%c2820: min = -7.481 -> 3.3036954
%       max = -7.477 -> 3.3342641
%                 => 0.9%
%
%MgIIk: min = -5.571 -> 2.6853444
%       max = -5.46  -> 3.4673685
%                 => 25% (28% in paper)
%
%c1506: min = -7.548 -> 2.831392
%       max = -7.510 -> 3.090295
%                 => 8.7% (10.4% in paper)
%
%alpha Cen B 
%MgIIk: 46%
%c1506: 51%
%
%alpha Cen A
%MgIIk: 13%
%c1506: 8%
Few other {Sun-like} stars have sufficient NUV monitoring data sets available to assess
their variability on activity cycle timescales.
{While \citet{ayers21} showcases NUV spectroscopic data for 
$\alpha$ Cen A+B and Procyon, he concludes that the continuum 
measurements are unreliable due to variable narrow-slit throughput
and thus does not assess NUV continuum variability.
FUV data for $\alpha$ Cen A reveals 1506\,\AA\ 
continuum variability over one cycle of $\approx$8\% and hence comparable to what is
seen for the Sun. Such results seem in tension with the study of
\citet{sofia89} which found through {\it International Ultraviolet Explorer}
({\it IUE}) 1715-1915\,\AA\ spectroscopic monitoring of $\alpha$ Cen A
$\sim$20\% continuum variability over a 3~year timescale (a small fraction of its $\sim$20~year
activity cycle).}
The reality of the {\it IUE} result remains to be verified, a task we leave
to a future effort.

Returning our attention to M67, we consider the Ca\,II activity monitoring
program of \citet{giampapa06} which {for a sample
of M67 Sun-like stars similar to that presented herein
uncovered interesting variability properties}.
They observed their targets over
1996-2002, a total duration comparable to the {\it GALEX} monitoring
reported here. They observe variability in nearly all their targets and in
some cases see patterns similar to activity cycles with periods $>$6~years.
M67 Sun-like star Ca\,II variability consistently has a higher amplitude than that
seen for the Sun, a result that would seem supportive of possible 0.3~magnitude
NUV variability in the {\it GALEX} sample
{as both should be due to magnetic activity processes and
if one activity indicator shows enhanced amplitudes relative to the Sun then it might be plausible for other activity indicators to show similar behavior}. We reiterate that we consider such
a result inconclusive, but certainly worthy of deeper investigation.

%Brasseur19: which results in pbol=0.132. All of the energies shown in the plots in this paper have been corrected to a bolometric energy value; multiplication by this value of pbol will give the radiated energy in the GALEX NUV bandpass.
% I've corrected all energy values to NUV bandpass only!
Our flare search results can be directly compared to the extensive assessment
of NUV flare events in the {\it Kepler}-{\it GALEX} survey conducted by
\citet{brasseur19} which featured a sample dominated by Sun-like stars.
Specifically, \citet{brasseur19} provide context around what ultraviolet flares
on such stars would look like (their Figures 3-5 
provided impetus for our flare search requirements), 
their distributions of durations and energies
(note that we correct all flare energy values shown in the \citealt{brasseur19}
figures to that radiated in the {\it GALEX} NUV bandpass only, achieved
by multiplying by a factor of 0.132 as described in their Section 3.2.3),
and the flare incidence rate as a function of flare integrated energy.
Our lightcurve sampling with 50~second time-bins and detection requirement
of two consecutive flaring data points effectively limits us
to detecting flares with duration of $\gtrsim$100~seconds
(which happens to coincide with the peak in the flare duration
distribution shown in Figure 11 of \citealt{brasseur19}).
Figure 13 of \citet{brasseur19} shows that flares with such durations
can have peak flare flux of anywhere from $\sim$1-1000$\times$ the stellar
quiescent NUV flux level. Such flares generally have NUV energies
between 10$^{32}$-10$^{36}$\,erg (Figure 16 of \citealt{brasseur19}
after correction).
Figure 14 of \citet{brasseur19} show their recovered flare incidence rate
as a function of integrated energy.

We seek to assess whether or not the M67 Sun-like star sample features similar
flare incidence rates as the {\it Kepler}-{\it GALEX} survey of \citet{brasseur19}.
For any individual M67 star on its own, with a typical total NUV monitoring time
of $\approx$8,000~seconds, it is not possible to broach the incidence rates
for even the weakest flares ($\sim$10$^{32}$\,erg) which are around 10$^{-5}$
flares star$^{-1}$ second$^{-1}$ (our flare rate limit in such a case would
be $\lesssim$10$^{-4}$ flares star$^{-1}$ second$^{-1}$).
However, under the assumption that the magnetic properties of all M67 Sun-like
stars are on average the same, we can make our assessment based on 
the aggregate time spent monitoring the cluster's stars. For such an assumption
we find that 4.716$\times$10$^{5}$ seconds of M67 Sun-like star NUV monitoring
were obtained by {\it GALEX} and our effective flare incidence rate limit
across NUV energies of 10$^{32}$-10$^{36}$\,erg is
$\lesssim$2.1$\times$10$^{-6}$ flares star$^{-1}$ second$^{-1}$
($\lesssim$1 NUV flare star$^{-1}$ week$^{-1}$).

Comparing our flare incidence rate limit 
to Figure 14 of \citet{brasseur19} shows agreement 
for flare NUV energies of $\gtrsim$10$^{33}$\,erg.
However, for flares with NUV energies between 10$^{32}$-10$^{33}$\,erg
our incidence rate limit should have resulted in the detection of a few to
several such flares according to the incidence rates of \citet{brasseur19}
which range between 2$\times$10$^{-6}$ to 10$^{-5}$ flares star$^{-1}$ second$^{-1}$.
We can think of two possible explanations for this tension:
1) our assumption that treats all M67 Sun-like stars in aggregate is unreasonable
{(the bifurcated FUV magnitude distribution in the bottom panel of Figure \ref{fig:nuvhr}
could be evidence for this as it may be suggestive of distinct activity properties within the
population)};
2) M67 stars are generally older than the flaring population displayed in
Figure 14 of \citet{brasseur19} and flare rates are depressed for older stars.
Age information is not presented in \citet{brasseur19} and age-adjacent
information like rotation periods are only reported for a small fraction of their
survey sample, thus we cannot conclusively explore the second explanation
although it seems consistent with optical flaring results presented
in \citet{davenport19}.
%???can I find a flare energy frequency distribution for the Sun for any wavelength?

The flares explored here are vastly more energetic than the typical
M-X class flares observed on the Sun (Figure 21 of \citealt{brasseur19} and
references therein). However, based on the work of \citet{tilley19} and
references therein, the derived flare rate limits 
suggest conditions that are unlikely to threaten the habitability of
any orbiting Earth-like planets, especially if associated energetic
particles do not routinely impact the planet (the likelihood of such impacts is much lower
for Earth-like planets that have orbital semi-major axes of $\approx$1\,AU
compared to the M-dwarf habitable-zone planets studied by \citealt{tilley19}).

%talk about UVEX and what I want it to do and what it is supposed to do
% age flare-rate dependence, lots of clusters in UVEX!
%what should UVEX do: long term monitoring with at least yearly visits to a variety of open clusters with <~1% precision in FUV and NUV per visit
%                     want to accrue lots of aggregate time for a variety of clusters with lightcurve sampling on the order of <~50 seconds (10 would be ideal); with ~50 stars per spectral type bin and a few visits per year of at least 3600 seconds duration this would produce 540,000 seconds of aggregate monitoring per year -- over 5 years would begin to get into the 10^-7 flares/sec domain; a visit every other month for 10 total years would broach 10^-8 flares/sec and most observed flare energies from Brasseur+'19
This work has demonstrated that {\it GALEX} capabilities and monitoring durations for
the M67 cluster were just short
of addressing the science goals of this investigation. UVEX is the planned
successor to {\it GALEX} and we briefly discuss how it could deliver
on assessing cluster ultraviolet variability properties;
we rely on parameters described in \citet{kulkarni21} and various updates
on the UVEX webpage at the time of writing this article.
UVEX is planned to have a prime mission lifetime of 2~years and
will survey the entire sky in two bandpasses that resemble the {\it GALEX}
NUV and FUV filters. The wide field-of-view (3.5$^{\circ}$) and finer
resolution than {\it GALEX} ($\approx$2$'$$'$ compared to $\approx$5$'$$'$) 
will make UVEX an efficient
cluster surveying machine. While UVEX is supposed to boast 50-100$\times$
better sensitivity than {\it GALEX} in a nominal survey field, it is not
yet clear what the photometric repeatability will be (recall that {\it GALEX}
was limited to 0.03~magnitudes repeatability in the NUV band). 
At present, the prime mission 
sky survey is slated to cover each point on the sky at a minimum of 10 times
with cadences ranging from 12~hours to 6~months. UVEX is expected to feature
an extended mission after the 2~year prime mission.

To expand upon the research presented herein UVEX must feature
the following characteristics.
Photometric repeatability between visits at the 
$\lesssim$1\% level is critical to detecting long-term activity 
cycle variability from Sun-like stars.
Clusters must be visited at least once per year over a 5-10 year baseline
to probe activity cycle durations comparable to the that of the Sun.
It is not clear how UVEX will read-out individual exposures or what the
minimum sampling time could be; flares would be best explored with sampling
times of $\lesssim$10~seconds.
To explore flare statistics in a conclusive manner it is essential to
accrue a substantial amount of total monitoring time for each cluster.
Assuming UVEX observes rich clusters featuring $\sim$50 stars per FGKM
spectral type bin and obtains 7200~seconds of total monitoring 
every year (e.g., 8 visits of 900~seconds duration each),
this would result in an aggregate of 360,000~seconds of monitoring
per year per spectral type bin. Maintaining this observing strategy
for a 5~year mission duration
would allow flare rates in the 10$^{-7}$ flares star$^{-1}$ second$^{-1}$
regime to be explored. Covering clusters with a wide variety of ages
would provide the most comprehensive age-ultraviolet flare rate
relation ever constructed; more impressive still is that it would be
partitioned into spectral type bins, thus opening up potent use cases
for understanding planetary atmosphere evolution and stability.

%% Also note that the akcnowlodgment environment does not support long amounts of text. If you have a lot of people and institutions to acknowledge, do not use this command. Instead, create a new \section{Acknowledgments}.
\begin{acknowledgments}
We thank C.\ Million for help with interpreting {\it GALEX} time. We
appreciate the referee doing a thorough reading of the paper
and providing thoughtful comments that helped improve it. This work
has been partially supported by
HST-AR-17036.001-A and NASA ADAP grant 80NSSC23K0468. 
This research has made use of the VizieR service.
This work has made use of data from the European Space Agency (ESA) mission
{\it Gaia} (\url{https://www.cosmos.esa.int/gaia}), processed by the {\it Gaia}
Data Processing and Analysis Consortium (DPAC,
\url{https://www.cosmos.esa.int/web/gaia/dpac/consortium}). Funding for the DPAC
has been provided by national institutions, in particular the institutions
participating in the {\it Gaia} Multilateral Agreement.
\end{acknowledgments}


\vspace{5mm}
\facilities{\it GALEX}


\software{gPhoton \citep{million16}
          }

}



\appendix

\section{Epochal and Flare Lightcurve Examples}
\label{app:A}

\begin{figure}[!b]
    \centering
    %trim={left bottom right top},clip
    \begin{minipage}{80mm}
     \includegraphics[width=80mm,trim={0.25in 2.9in 0.75in 3in},clip]{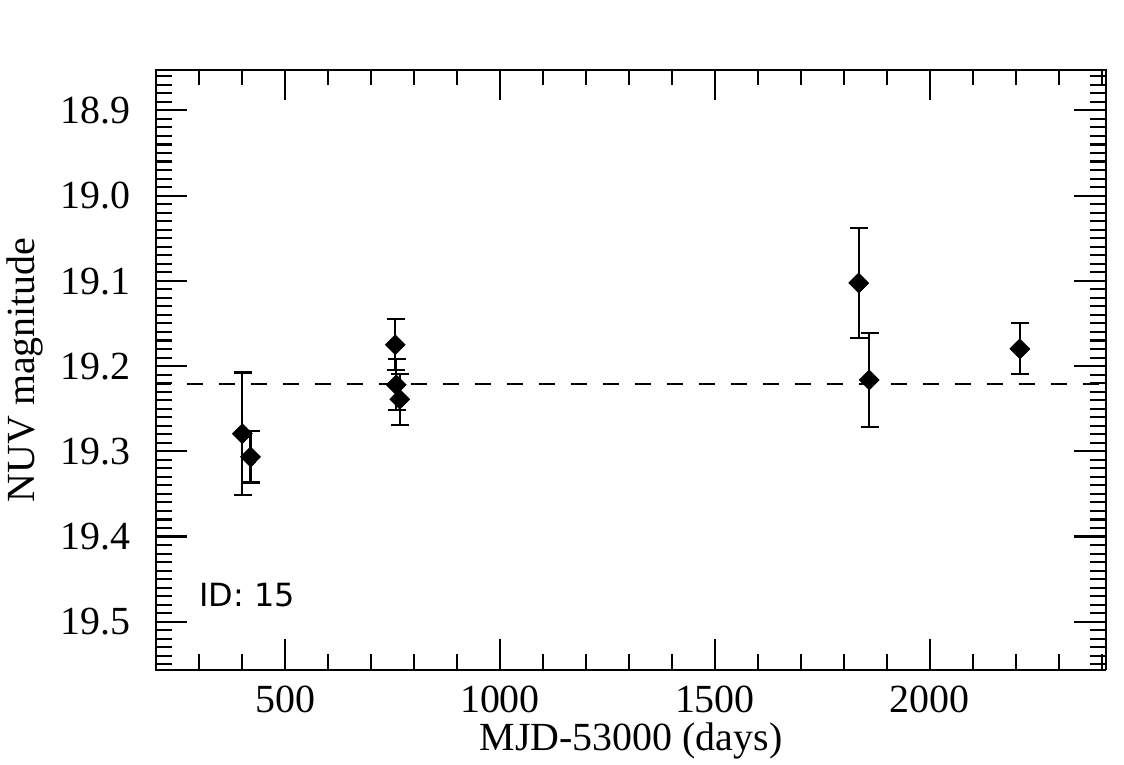}
    \end{minipage}
    \begin{minipage}{80mm}
     \includegraphics[width=80mm,trim={0.25in 2.9in 0.75in 3in},clip]{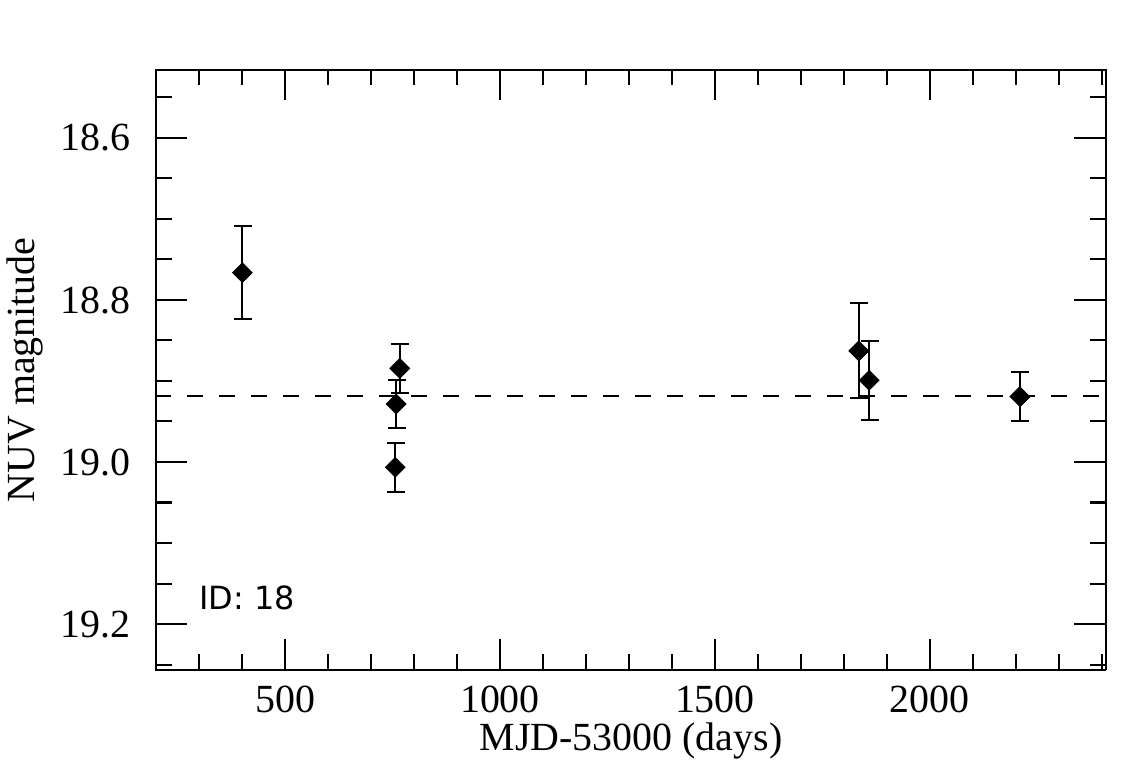}
    \end{minipage}
    \\*
    \begin{minipage}{80mm}
     \includegraphics[width=80mm,trim={0.25in 2.9in 0.75in 3in},clip]{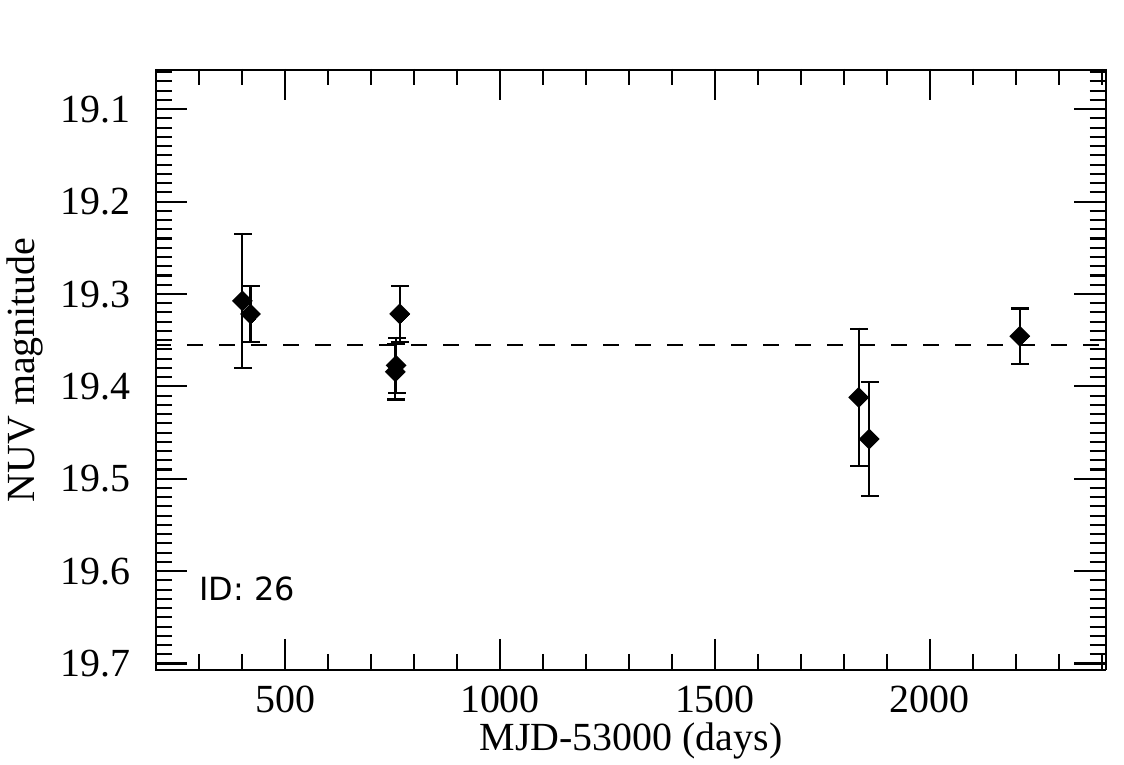}
    \end{minipage}
    \begin{minipage}{80mm}
     \includegraphics[width=80mm,trim={0.25in 2.9in 0.75in 3in},clip]{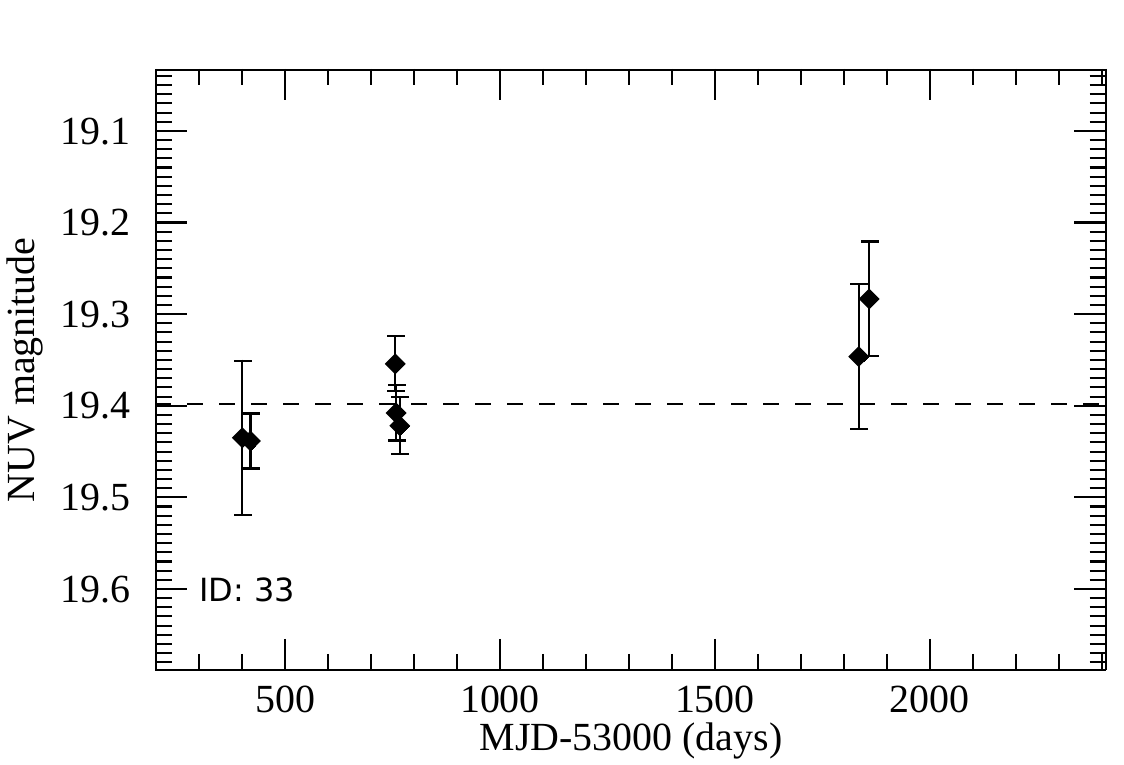}
    \end{minipage}
    \\*
    \begin{minipage}{80mm}
     \includegraphics[width=80mm,trim={0.25in 2.9in 0.75in 3in},clip]{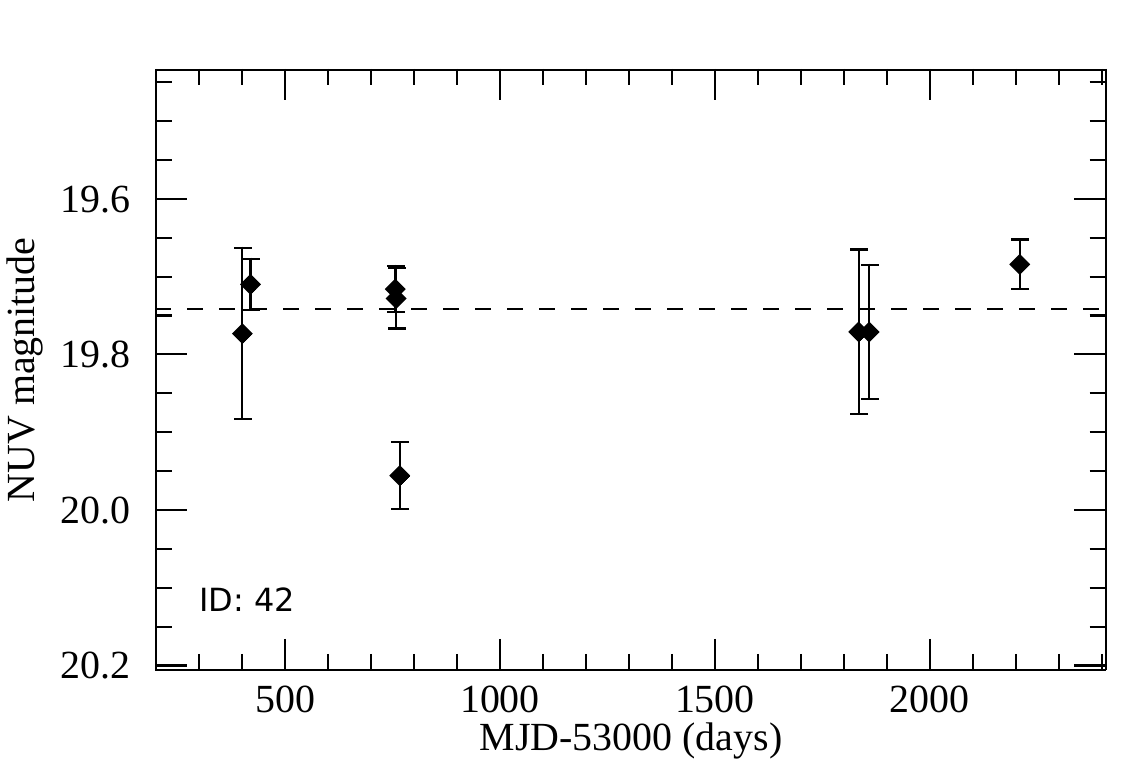}
    \end{minipage}
    \begin{minipage}{80mm}
     \includegraphics[width=80mm,trim={0.25in 2.9in 0.75in 3in},clip]{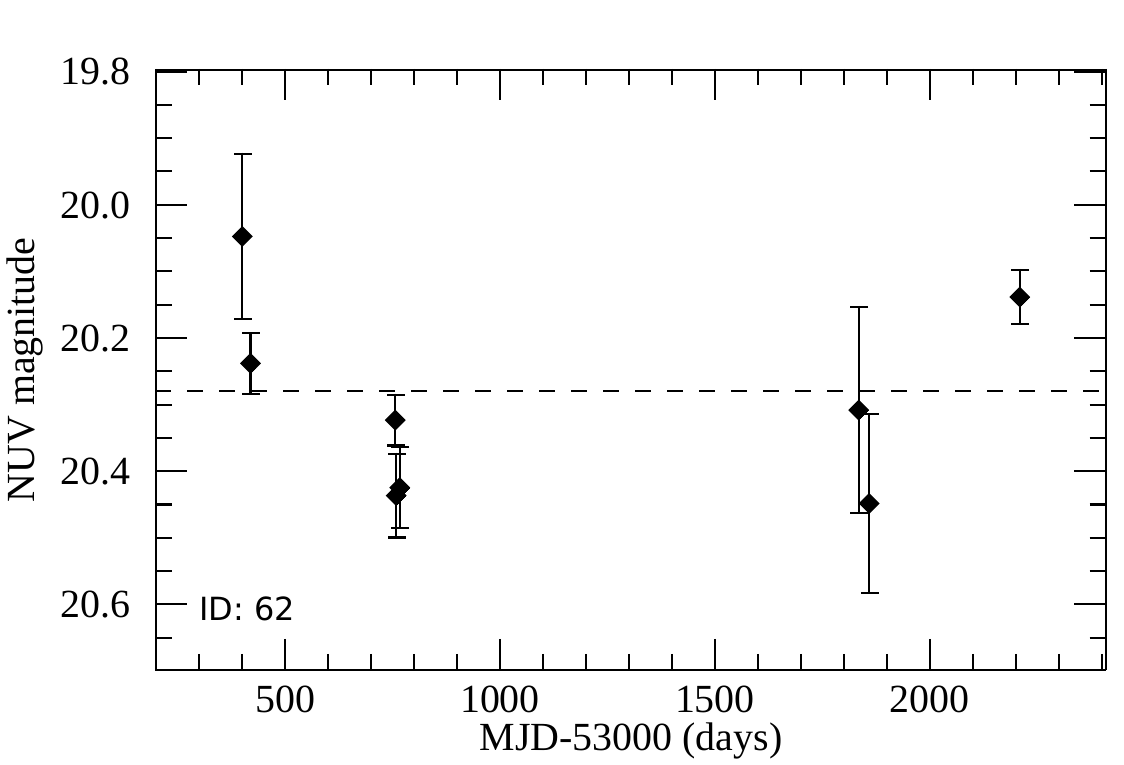}
    \end{minipage}
    \caption{Example epochal lightcurves; NUV magnitudes do not have aperture
    corrections applied. The dashed line in each figure panel shows the 
    weighted mean NUV magnitude calculated from all epochs.
    Star 42 is an example of a star
    with $>$5$\sigma$ difference between two epochs, in this case caused by
    one data point. One other star shares such characteristics. Stars 15, 18,
    and 62
    have 3-5$\sigma$ differences between two epochs and also appear to have
    weak evidence for long-term variability like what one would expect for
    an activity cycle with period $\gtrsim$5~years. Stars 26 and 33 have 2-3$\sigma$
    evidence for long-term variability from what could be an activity cycle.}
    \label{fig:epcex}
\end{figure}

\begin{figure}[!h]
    \centering
    %trim={left bottom right top},clip
    \begin{minipage}{80mm}
     \includegraphics[width=80mm,trim={0.25in 2.9in 0.75in 3in},clip]{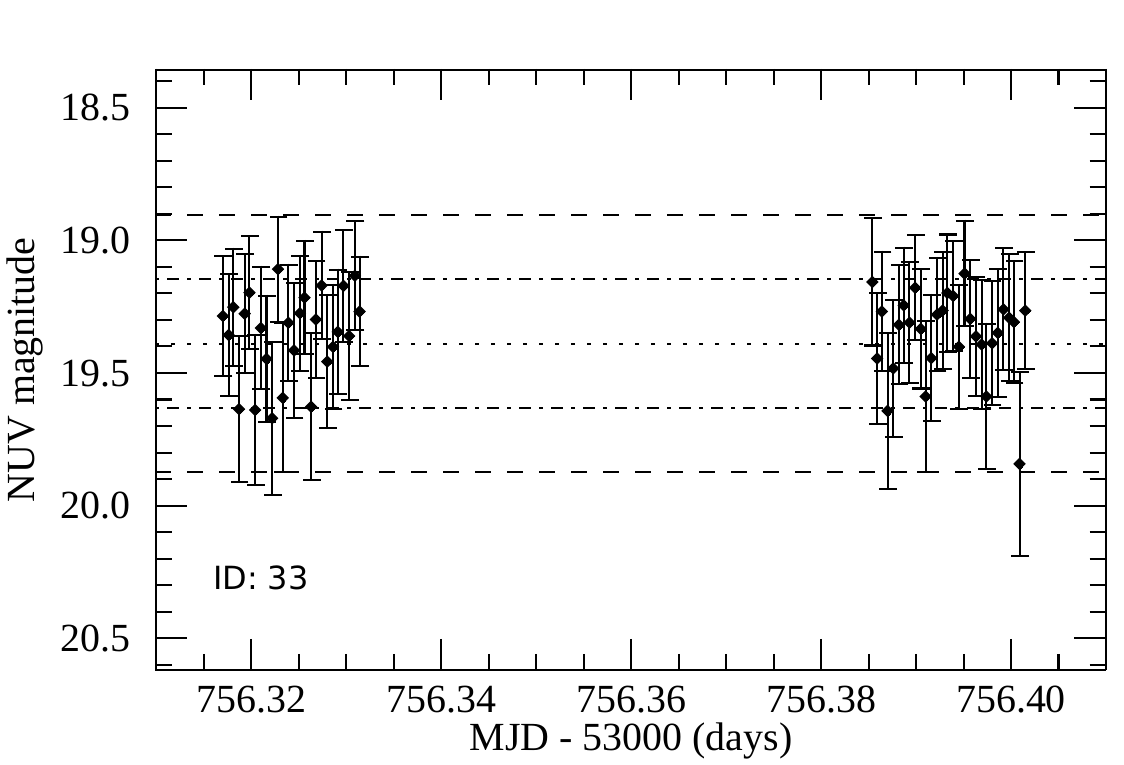}
    \end{minipage}
    \begin{minipage}{80mm}
     \includegraphics[width=80mm,trim={0.25in 2.9in 0.75in 3in},clip]{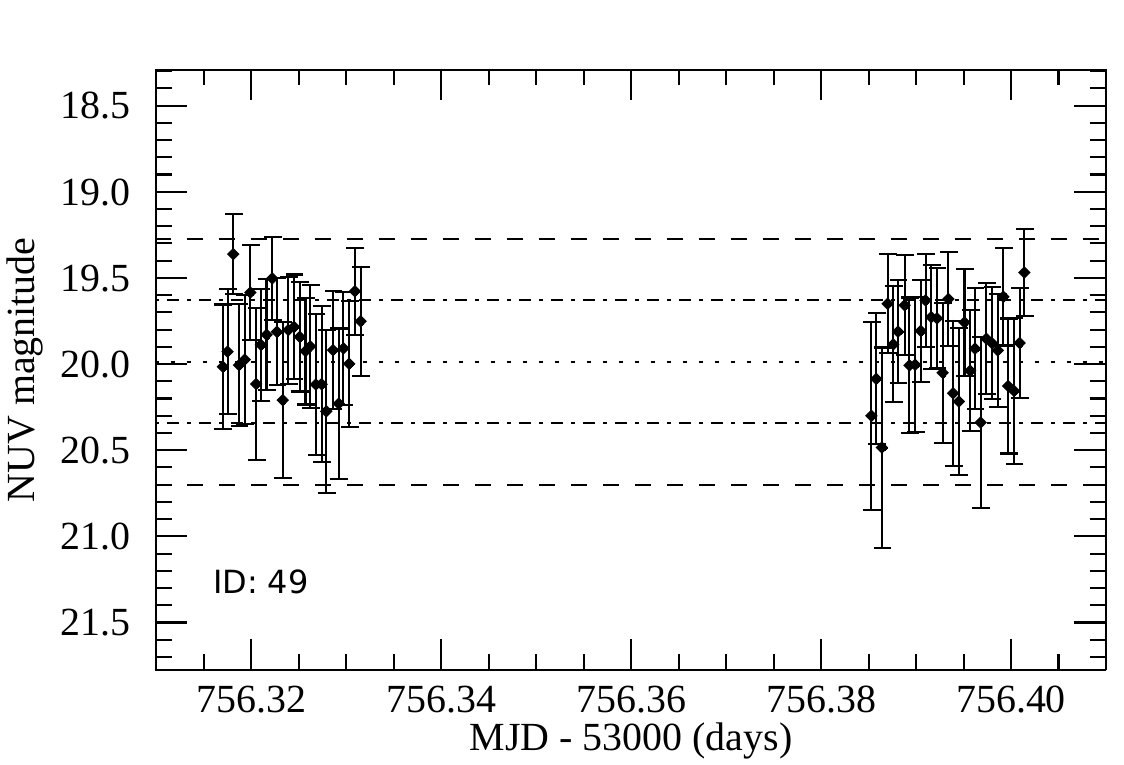}
    \end{minipage}
    \caption{Example flare lightcurves zoomed in on Epoch 3 time regions
    (see Table \ref{tab:epochs}); NUV magnitudes do not have aperture
    corrections applied. The dotted line in each figure panel shows the 
    median NUV magnitude calculated from all flare lightcurve samples.
    The dash-dotted curves show the median magnitude value $\pm$1$\times$ the
    median uncertainty value from all samples and the dashed curve $\pm$2$\times$
    this value.}
    \label{fig:lcex}
\end{figure}

\vskip -0.03in
In this Appendix we show example figures of stars whose epochal
lightcurves either appear to show variability or had $>$3$\sigma$
magnitude differences between epochs (Figure \ref{fig:epcex}). 
Two example flare lightcurve segments are also shown
(Figure \ref{fig:lcex}).

\clearpage

\startlongtable
\begin{deluxetable}{lcccccccccc}
\label{tab:phot}
\tabletypesize{\small}
\tablewidth{0pt} 
\tablecaption{{\it GALEX} Photometry}
\tablehead{ \colhead{ID} &
            \colhead{Ap.} &
            \colhead{Annulus} &
            \colhead{NUV Ap.} &
            \colhead{FUV Ap.} &
            \colhead{NUV Epoch} &
            \colhead{NUV} &
            \colhead{NUV Mag.} &
            \colhead{FUV Epoch} &
            \colhead{FUV} &
            \colhead{FUV Mag.} \\
            \colhead{} &
            \colhead{Rad.\ ($'$$'$)} &
            \colhead{Rad.\ ($'$$'$)} &
            \colhead{Corr.\ (mag)} &
            \colhead{Corr.\ (mag)} &
            \colhead{(MJD)} &
            \colhead{Mag.} &
            \colhead{Error} &
            \colhead{(MJD)} &
            \colhead{Mag.} &
            \colhead{Error} \\
}
\startdata
  \multirow{8}{*}{1} &         \multirow{8}{*}{12.8 } &              \multirow{8}{*}{16.0,22.0} &    \multirow{8}{*}{0.090} &    \multirow{8}{*}{0.090} &   53400.5808 & 18.674 & 0.054 &    \multirow{8}{*}{53724.6400} &   \multirow{8}{*}{23.792} &     \multirow{8}{*}{$-$} \\
                     &                                &                                         &                           &                           &   53419.5908 & 18.588 & 0.030 &                                &                           &                           \\
                     &                                &                                         &                           &                           &   53756.3515 & 18.571 & 0.030 &                                &                           &                           \\
                     &                                &                                         &                           &                           &   53758.3800 & 18.570 & 0.030 &                                &                           &                           \\
                     &                                &                                         &                           &                           &   53766.7386 & 18.576 & 0.030 &                                &                           &                           \\
                     &                                &                                         &                           &                           &   54834.4817 & 18.559 & 0.049 &                                &                           &                           \\
                     &                                &                                         &                           &                           &   54858.8591 & 18.606 & 0.040 &                                &                           &                           \\
\enddata
\tablecomments{ID is the source identification number given in Table \ref{tab:targs}. 
Ap.\ Rad.\ is the aperture radius used to extract photometry for both bands. 
Annulus Rad.\ is the inner, outer radii respectively used to define the background annulus region.
NUV and FUV Ap.\ Corr.\ are the aperture corrections for each band; to apply these corrections, subtract them from the reported photometry in the table.
NUV Epoch is the Modified Julian Date of the photometric value reported for each row. NUV Mag.\ and Mag.\ Error are the extracted AB magnitude and associated uncertainty respectively for the corresponding epoch.
FUV Epoch reports the average Modified Julian Date for each source after integrating all available FUV data. FUV Mag.\ is the corresponding extracted AB magnitude. FUV Mag.\ Error reports the uncertainty when a source detection is made (see Section \ref{sec:methods}), otherwise a ``$-$'' indicates the FUV Mag.\ is an upper limit.}
\end{deluxetable}



\end{document}